# Applications of Knot Theory for the Improvement of the AlphaFold Protein Database

Pranshu Jahagirdar

**Abstract**

AlphaFold, a groundbreaking protein prediction model, has revolutionized protein structure prediction, populating the AlphaFold Protein Database (AFDB) with millions of predicted structures. However, AlphaFold's accuracy in predicting proteins with intricate topologies, such as knots, remains a concern. This study investigates AlphaFold's performance in predicting knotted proteins and explores potential solutions to enhance the AFDB's reliability. Forty-five experimentally verified knotted protein structures from the KnotProt database were compared to their AlphaFold-generated counterparts. Knot analysis was performed using PyKnot, a PyMOL plugin, employing both Gauss codes and Alexander-Briggs knot notations. Results showed 95.6% accuracy in predicting the general shape of knots using Alexander-Briggs notation. However, Gauss code analysis revealed a 55.6% discrepancy, indicating AlphaFold's limitations in accurately representing the intricate orientation and directionality of knots. This suggests potential inaccuracies in a significant portion of the AFDB's knotted protein structures.

Applications of Knot Theory for the Improvement of the AlphaFold Protein Database

The study underscores the need for improved knot representation in AlphaFold and proposes potential solutions, including transitioning to a single-module design or removing incorrectly predicted structures from the AFDB. These findings highlight the importance of continuous refinement for AI-based protein structure prediction tools to ensure the accuracy and reliability of protein databases for research and drug development.

## Introduction

**Background**

Proteins are macromolecules that are found almost everywhere within the human body and perform many critical functions such as catalysis, structure formation, nutrient transport, and body defense. Proteins, which are made up of twenty standard amino acids, must take a specific three-dimensional shape to function correctly. The amino acids come together to form such a shape through a process called protein folding. Protein folding is the intricate process by which a linear chain of amino acids, the building blocks of proteins, assumes its three-dimensional, functional structure. The sequence of amino acids dictates how a protein folds and the resulting entanglement determines its ability to interact with other molecules.

**Rationale**

It is possible, however, for amino acid chains to not always fold correctly. Protein misfolding is believed to be the cause of many neurodegenerative diseases, called amyloid diseases, such as Parkinson's disease, cystic fibrosis, and Alzheimer's. Currently, medical professionals can restrict the damage caused by misfolded proteins through the use of molecule inhibitors. These molecule inhibitors "target proteins' function by binding to the 'pocket' on their surface" (Liu et al., 2020). This binding interaction interferes with the normal activity of the protein, disrupting its function. In the context of protein misfolding and diseases, such as neurodegenerative disorders, molecule inhibitors can help mitigate the damage caused by misfolded proteins. Not knowing the structure of a target protein can make it difficult to develop a suitable molecule inhibitor for that specific protein. So far, only about 200,000 protein structures have been experimentally analyzed and saved in the Protein Data Bank. Protein structures have long been modeled through the use of X-ray crystallography and NMR spectroscopy though these methods are tedious and time-consuming as they require manual labor as well as specialized equipment. These methods face limitations in resolving complex protein structures and can be costly. Fortunately, artificial intelligence combined with the field of mathematics called knot theory can fill in this knowledge gap significantly faster by predicting how proteins fold, what structure they may become after folding, and filtering out generated structures that could not exist in reality.

## Review of Literature

Many AI models have already been developed for protein structure prediction, such as AlphaFold and RoseTTAFold. This paper focuses more specifically on AlphaFold 2, an improved version of AlphaFold that generates the atomic coordinates within a protein as well as a confidence score based on the lDDT-Cα (pLDDT) metric for every residue within the generated protein model on a scale from 1-100. As of September 2023, AlphaFold 2 has added 214,000,000 protein structures to its own AlphaFold Database (Varadi et al., 2023).

**Problem Statement**

Applications of Knot Theory for the Improvement of the AlphaFold Protein Database

While AlphaFold is generally very accurate, it struggles to predict protein structures with non-trivial topologies such as knots, slip-knots, and links. According to Niemyska et al. (2022), "Currently around 2% of solved protein structures from the PDB are considered to contain non-trivial topologies: knots, slipknots, or links". These rare topologies can be analyzed and represented in mathematical notation through the utilization of knot theory. Even though a knot is defined by knot theory as a "closed curve," Millett et al. (2013) point out that knot theory techniques can still be utilized to examine knots in proteins by joining their N-terminus and C-terminus ends.

**Figure 1**

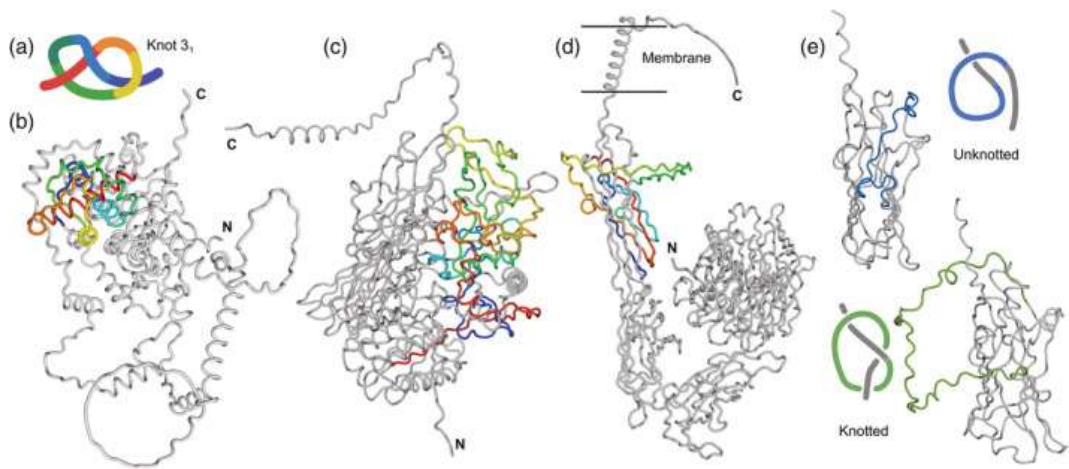

*Figure 1: The most common knot type found in proteins is written as $3_1$ in the Alexander–Briggs notation (Perlinska et al., 2023)*

One reason why AlphaFold struggles with non-trivial topologies is due to its dual module design. As explained by Jumper et al. (2021), AlphaFold 2 is composed of two main modules: the Evoformer module, which generates an abstract representation of the protein structure, and the Structure module, which generates a 3D model by rotating and translating all of the residues within the protein. A two-module system allows AlphaFold to run much more efficiently as each module can perform a specialized task rather than a single module taking a more generalized approach. To simplify calculations, however, Stasiak et al. (2023) point out that the residues are allowed to move freely while being adjusted by the structure module, which may cause AlphaFold to violate certain peptide bond geometries and generate protein structures that could not be created during the real protein folding process. Another reason why AlphaFold may struggle with knots is because of its reliance on homologous structure data. As described by Evans et al. (2022), AlphaFold utilizes homologous structures to generate the structure of the target protein by taking known homologous complexes and using them as a template. This TBM approach is beneficial when the target protein has a close homolog with an experimentally solved structure in the Protein Data Bank. However, being evolutionary related to another protein does not completely determine whether a protein will have a knot or not. This can be seen in a study conducted by Perlinska et al. (2023), where AlphaFold predicted some of the integrin structures to contain the $3_1$ knot even though the homolog set used had a mix of knotted and unknotted structures. The authors do point out that experimental analysis is still required in order to verify whether AlphaFold has correctly predicted these structures. The geometric violations made by AlphaFold may have not been caught by the aforementioned pLDDT metric which can give high scores to proteins with incorrect non-trivial topology because of its inability to account for knots. As indicated by Dabrowski-Tumanski et al. (2023), the pLDDT confidence metric looks at residues individually without taking into account the protein backbone in its entirety.

      DeepMind, the organization behind AlphaFold, and the European Bioinformatics Institute (EMBL-EBI) have used AlphaFold2 to generate predicted structures for millions of proteins across various organisms. These predicted protein structures, even with potentially mispredicted knotted proteins, are then compiled and made publicly available through the AlphaFold Protein Database (AFDB) to researchers globally on AlphaFold's website. As it stands, there are no apparent statistics on the number of incorrectly knotted proteins in the AFDB

Applications of Knot Theory for the Improvement of the AlphaFold Protein Database

made by AlphaFold and the severity of potential mispredictions made by it. A proposed solution (and the inspiration for the methodologies for this study) for determining the scale of possible errors within the database by Dabrowski-Tumanski et al. was obtaining knot invariants from the KnotProt web server and using a Python script named Topoly to perform knot analysis in order to verify whether AlphaFold had represented knots correctly in its generated protein model. These predicted methods would allow for an estimation of how many incorrect knotted models are in the AFDB and would allow for the identification of specific structures that are incorrectly predicted on the Database during the course of the study which could then be requested to be removed by contacting EMBL-EBI or DeepMind.

## Methodology

According to Dabrowski-Tumanski et al, a comparison of experimentally derived knotted proteins with their AlphaFold-generated counterparts is necessary to determine AlphaFold's prediction accuracy of knotted proteins. Such a comparison would assist in determining the extent to which AlphaFold's database is inaccurate, as well as determining certain protein structures that are incorrect and available to the public on the AFDB, which could then be requested to be removed, thereby improving the quality of the database.

**Experimental Knotted Proteins**

Following Dabrowski-Tumanski's suggestion, the KnotProt database, a database of proteins with knots, slipknots, and knotoids, was used to find experimentally determined models of knotted proteins. Specifically, 45 knotted proteins were downloaded from this database. Additionally, only proteins with whole knots were chosen, and those with broken chains or slip knots were not included. This is due to the fact that the knot analysis tool utilized in this study could not handle these specific topological entanglements and that any analysis done on an artificial or broken knot could reduce the quality of conclusions derived from the said analysis.

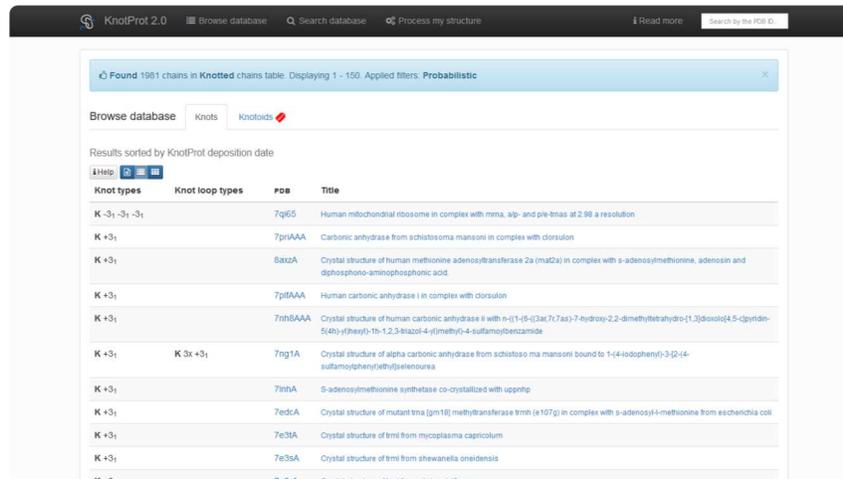

*Figure 2: The KnotProt database*

**Protein Structure Search**

To obtain the AI-generated counterparts of the selected KnotProt proteins, this study utilized Foldseek, a search engine developed to help researchers find proteins across multiple databases. Foldseek offers the functionality of uploading PDB (Protein Data Bank File) files to their search server which allows for the lookup of the previously downloaded KnotProt proteins across a variety of protein databases. Out of the list of possible protein matches within the search results, models that were stored on the AlphaFold Database specifically were downloaded. All of the models searched had hundreds of similar structures stored on the AlphaFold Database. In order to choose the protein structure that is most likely to be modeled after the same protein as the KnotProt protein, models with matching parent organisms were chosen. This reduces the likelihood that the compared KnotProt and AlphaFold structures were not modeled after the same target protein which would cause any results to appear as though AlphaFold was severely mispredicting knotted protein structures when that may not actually be the case. The search results consist of links to the protein's webpage on the AlphaFold Database.

      Utilizing a third-party protein search tool was critical for the efficiency aspect of this study, even though the AlphaFold database has its own native sequence searching capabilities built into the website. The AlphaFold database provides a sequence search tool that allows users to input a protein sequence and retrieve a result of AlphaFold structure predictions with similar sequences from its database. Specifically, the sequence search tool compares the input protein sequence against the sequences of the predicted AlphaFold models in the database and returns a list of matches ranked by sequence similarity. However, AlphaFold's sequence search can be quite slow due to several factors. The first is that AlphaFold's sequence search relies on building multiple sequence alignments (MSAs) for each search against the database, which can be time-consuming, especially for longer sequences or those without close homologs (Van Kempen et al.). The sequence search also lacks the feature of pre-computed indexes, and unlike tools such as Foldseek that use pre-computed database indexing to enable rapid searches, AlphaFold's native search implementation does not leverage such techniques, leading to slower searches. The speed advantage of using Foldseek over AlphaFold's native search can be around 4 orders of magnitude (Van Kempen et al.).

Applications of Knot Theory for the Improvement of the AlphaFold Protein Database

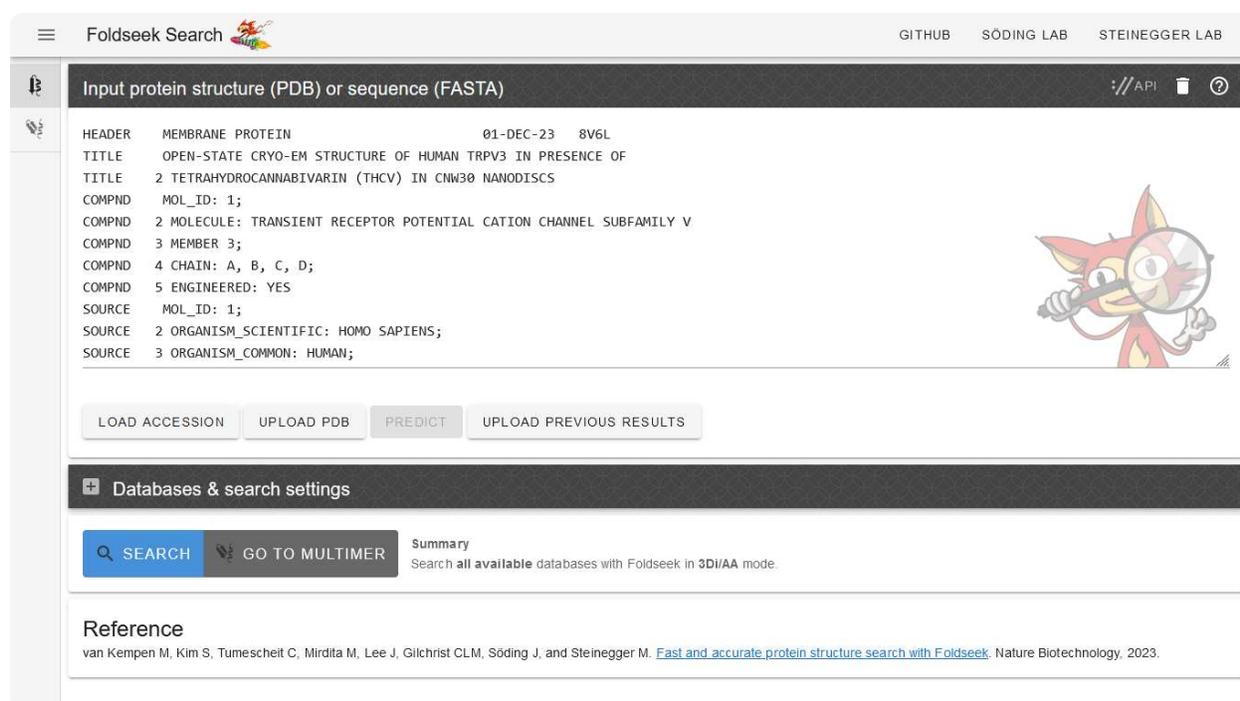

Figure 3: Foldseek Search Server

**Knot Analysis**

The initial method for performing knot analysis was to use a scripting approach using Topoly, a polymer topology analysis tool, in order to compare the qualities of the knots, if any are present, within the AlphaFold predicted models against the KnotProt models, as recommended by Dabrowski-Tumanski et al. While Topoly supports a wider range of knot notations, it lacks protein visualization capabilities because it is restricted to the command line and produces numerous import errors during use. The alternative to Topoly used in this study was PyKnot, a PyMOL plugin that has the capability of performing knot invariant analysis. The advantage that PyKnot has over Topoly is its ability to provide visual knot analysis using PyMOL, a molecular visualization system. PyKnot is also quite computationally efficient, allowing for faster analysis on local hardware and removing the need for any cloud computing power.

Applications of Knot Theory for the Improvement of the AlphaFold Protein Database

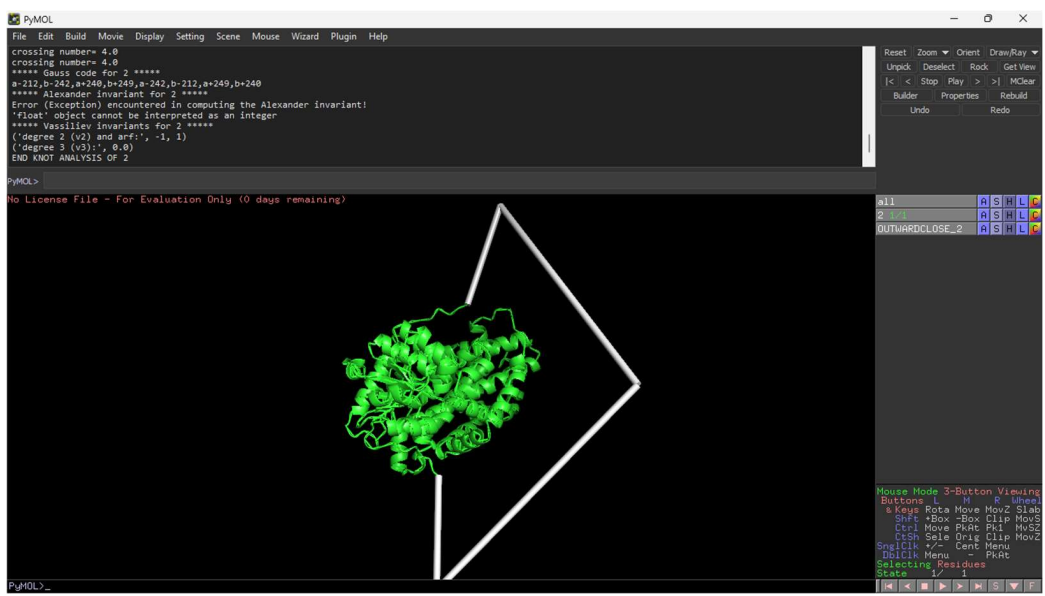

*Figure 4: The PyMOL program with an imported protein model, having knot analysis performed on it*

To begin knot analysis, a KnotProt protein and its corresponding AlphaFold predicted version were imported into PyMOL, and PyKnot analysis was run on both proteins. The PyKnot program then proceeds to print the Gauss code and Alexander-Briggs knots invariants in the PyMOL terminal both of which are different notations for representing knots in mathematical format. By recording the knots within both proteins in mathematical notation, the ability to more accurately compare them and determine how different one knot is from the other is gained. In the initial stages of the study, the only knot notation that could be recorded was the Gauss code as the PyKnot plugin was not completely compatible with the PyMOL visualization system as it was written in Python 2.0 whereas PyMOL's latest version is written in Python 3.1. This version difference would cause the error `'float' object cannot be interpreted as an integer` in the command line output dialogue for the Alexander invariant part of the knot analysis result and also cause the Vassilev invariant to not be able to display any Alexander-Briggs notations. Fortunately, the PyKnot program's source code is freely available on GitHub allowing for the opportunity to fix this error without requiring assistance from the original developers of the

Applications of Knot Theory for the Improvement of the AlphaFold Protein Database

program. The float point error was caused by a simple division error where the result needed to be a whole number while in reality, it was giving a float value. This was easily rectifiable as it only required all true division operators (/) to be converted to floor division operators (//) which was done through the use of Visual Studio Code. The absence of any outputs by the Vassilev invariant calculator was caused by misplaced paratheses within the code which was also a very simple fix. By doing so, the Vassilev invariant calculator was now able to analyze knots and give the Alexander-Briggs notations of the analyzed knot allowing for 2 different knot notations to be recorded in this study, thereby potentially bringing new potential insights.

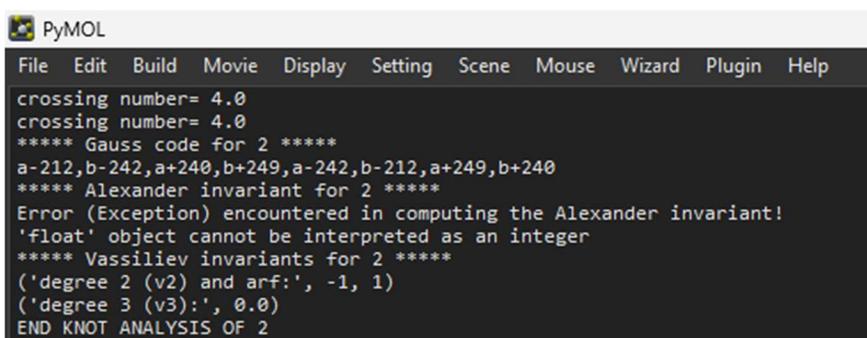

*Figure 5: The PyMOL terminal depicting errors in the knot analysis output*

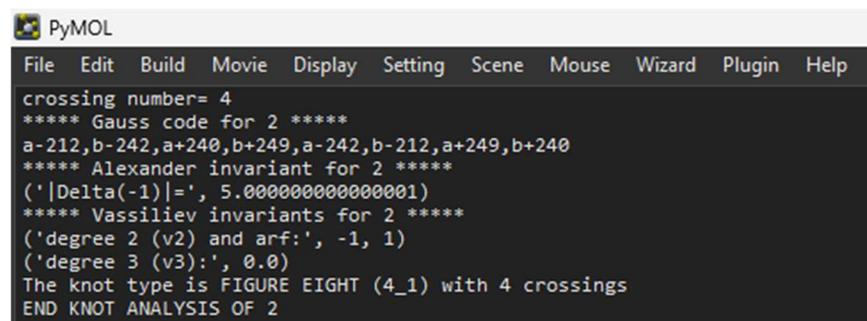

*Figure 6: The PyMOL terminal shows appropriate knot analysis results without errors*

**Data Collection and Analysis**

After analyzing the knots within both proteins, the Gauss codes and Alexander-Briggs knots were both recorded on a spreadsheet. Two separate pages were created for organizing each

Applications of Knot Theory for the Improvement of the AlphaFold Protein Database

knot notation type. Protein structures were stored in pairs allowing for easy comparison between the knots within the corresponding AI generated and experimentally derived models. After listing the knot type for each protein structure, it then becomes necessary to determine whether the knot within each model matches with the other. In the context of Gauss codes, knots can be determined to be the same if both contain the same sequence and number of under or over crossings. The Gauss code encodes the over or under-crossing pattern along the knot chain using + or – signs meaning that the matching order of the + or – signs would indicate that the knot is the same. Additionally, Gauss codes also contain two different letters in the knot sequence paired with each over or under-crossing to indicate which strand makes up that specific knot projection within the knot sequence which needs to be the same across both Gauss codes for the knots to be considered matching. The Gauss code also comes with a sequence of numbers that represents which crossing is being referred to in the Gauss code sequence. This study's main goal is to determine whether knots within the AI-generated model are representative of the knot in the experimental protein which means that this study will look at the specific orientations and direction of a knot within a protein meaning that the numeric values in the gauss codes will not be used for comparison as that would entail the AI-generated structure to have identical crossings which are unrealistic as protein structures that are modeled after the same protein will naturally always have slight differences in their exact crossing locations along a chain of residues.

Table 1: Comparison of Gauss codes

| PDB ID | KnotProt Protein Gauss Code | AlphaFold Protein Gauss Code | Match of Gauss Codes |
|---|---|---|---|
| 1uak | b-40,a-51,b-70,a-40,b-51,a-70 | b-36,a-37,b-44,a-36,b-37,a-44 | Yes |
| 1fug | a-6,b-14,a-86,b-6,a-14,b-86 | b-8,a-5,b-35,a-8,b-5,a-35 | No |

Applications of Knot Theory for the Improvement of the AlphaFold Protein Database

The second type of data collected was the Alexander-Briggs knots, which is a different notation for recording knots in a mathematical format that classifies knots based on their minimum crossing number. Compared to the Gauss code which records the exact sequence of over or under crossings along a knot, the Alexander-Briggs notation, on the other hand, offers a high-level overview of the general shape of the knot instead. This notation consists of two numbers: the first being the minimum crossing number and the second being a subscript number that distinguishes different knots with the same minimum crossing number. It should also be noted that this format does not include any extremely specific details about the knots being represented meaning that no parts of the notation will be excluded during analysis as was done with the Gauss codes with their crossing locations.

Table 2: Comparison of Alexander-Briggs Knots

| PDB ID | KnotProt Protein Alexander-Briggs Knot | AlphaFold Protein Alexander-Briggs Knot | Match of Alexander-Briggs Knots |
|---|---|---|---|
| 1uak | 3-1 | 3-1 | Yes |
| 2yfk | 3-1 | Unknotted | No |

## Conclusions

In this study, 95.6% of all the AI-modeled structures had matching Alexander-Briggs knots when compared to their experimentally derived counterparts. As mentioned previously, this specific type of notation describes a knot on a high level meaning that a 95.6% match rate indicates that AlphaFold can accurately represent the general shape of a knot within a predicted knotted structure. It should be noted that all of the AI-generated structures that did not have matching Alexander-Briggs knots were considered by PyKnot to be unknotted while their experimental counterparts were considered knotted. The two structures out of the total 90 structures compared did not have very complicated knots as they were both trefoils (3-1 knots in Alexander-Briggs notation) indicating that the pLDDT metric used in the prediction of these two specific knotted proteins showed a high level of confidence even though the predicted model lacked a knot highlighting the importance for having a metric that takes knots within proteins into account as errors such as this could continue to occur.

   A high level of matching accuracy was not the case for the comparison of Gauss codes, however. 55.6% of the AI-modeled proteins' Gauss codes did not match the Gauss codes of the KnotProt proteins, indicating a significant deviation from the experimentally verified structures. Since Gauss codes offer a much more descriptive representation of a knot in mathematical notation compared to the Alexander-Briggs invariant, it can be understood that the high number of non-matching Gauss codes indicates that while AlphaFold can represent the general shape of a knot within a protein fairly accurately, it cannot represent the directionality and orientation of a knot within a knotted protein accurately, meaning specific details of knots are not being correctly predicted. The orientation of a knot within a protein is extremely critical as it affects the folding pathway and kinetics of the protein. Proteins with different knot orientations may have different folding rates and mechanisms to form the knotted structure which could turn out to be very significant differences when conducting certain medical trials for potential treatments.

**Implications**

   While AlphaFold has not had any serious problems for proteins with "normal" topologies, proteins with more complicated structures are being mis-modeled, which could seriously impact the quality of research done by those who rely on the AFDB. Based on the percentage of mispredicted knotted protein structures in this study, 188,320 to as many as 2,379,680 proteins in the AlphaFold Database have been incorrectly modeled. This statistical range was calculated by multiplying the percentage of mispredicted knotted proteins (for the Alexander-Briggs Knots as well as the Gauss codes) by 0.02 (since 2% of the AlphaFold Database is made up of knotted proteins) and then further multiplied by 214,000,000 (the total number of protein structures in the database). While this may be a small percentage of the AFDB's 214 million entries, any researcher who uses a mispredicted knotted protein model for drug development may gain an incorrect understanding of a target protein's structure and thereby its functions which could seriously hinder the development of treatments such as molecule inhibitors. Developing a protein-targeting drug using a mismodeled structure could result in wasted time, money, and other resources, as the drug may fail in later clinical trials due to issues related to the incorrect structure.

**Solutions**

      A potential solution to AlphaFold's mispredicting of knotted proteins could be converting the model to a single module design. As previously indicated, AlphaFold uses a two-module design that allows residues to move freely during the prediction process. While this does reduce computer power, this could be the primary cause as to why AlphaFold misrepresents knots within proteins since allowing residues to move about freely causes AlphaFold to construct physically impossible knots. A single module design that restricts the movement of residues would make the predictions more realistic but could come at the cost of computational performance. To have the best of both worlds, DeepMind, the organization behind AlphaFold, could utilize the single module model only if it is extremely likely that the target protein is

knotted by taking a look at its homologs so that researchers would not have to unnecessarily use the more inefficient model for potentially knotted proteins. Developing such a module would certainly be a difficult and expensive task though this study provides a rationale for the creation of such a model by identifying the scale of the errors caused by AlphaFold's current model which are significant especially for the case of drug discovery. If the protein folding community and DeepMind desire to take the greatest measures possible to offer researchers the most accurate data on protein structures possible, it would be well worth the cost to develop the single-module AlphaFold Model.

   Throughout this study, 25 proteins predicted by AlphaFold were confirmed to be incorrect after comparing them to their experimental KnotProt counterparts. To improve some of the current entries within the AFDB directly, those 25 protein structures could be requested to be removed by contacting the authorities at EMBL-EBI or DeepMind, as previously mentioned. This solution would be effective on a much smaller scale compared to having a separate model to handle non-trivial topological cases but would still lead to a somewhat improvement of the AFDB. If the methodologies of this study were to be performed on a greater scale, around 200,000 protein structures which are incorrectly predicted could be taken down as there are 200,000 experimentally derived protein models to use for evaluation of AI models.

Applications of Knot Theory for the Improvement of the AlphaFold Protein Database

Applications of Knot Theory for the Improvement of the AlphaFold Protein Database

Applications of Knot Theory for the Improvement of the AlphaFold Protein Database

**Appendix**

(Contained on the next page)

# Applications of Knot Theory for the Improvement of the AlphaFold Protein Database

Table 3: Comparison of Gauss Codes (Full Data)

| PDB ID | KnotProt Knot | AlphaFold Knot | Knot Match |
|---|---|---|---|
| 7qi6 | b+15,a+116,b+144,a+15,b+116,a+144 | b+48,a+85,b+153,a+48,b+85,a+153 | Yes |
| 6km1 | a-11,b-13,a-31,b-11,a-13,b-31 | b-26,a-27,b-31,a-26,b-27,a-31 | No |
| 6ptx | a+3,b+4,a-49,b-77,a+4,b+3,a-77,b-49 | b-123,a+127,b+130,a-123,b-333,a+130,b+127,a-333 | No |
| 1ipa | a-78,b-89,a-91,b-78,a-89,b-91 | a-99,b-100,a-106,b-99,a-100,b-106 | Yes |
| 6ux1 | a-15,b-17,a-35,b-15,a-17,b-35 | b-26,a-27,b-31,a-26,b-27,a-31 | No |
| 3hkn | b-13,a-29,b-105,a-13,b-29,a-105 | b-26,a-27,b-31,a-26,b-27,a-31 | Yes |
| 4twm | b-11,a-17,b-20,a-11,b-17,a-20 | b-28,a-37,b-41,a-28,b-37,a-41 | Yes |
| 5vik | b+11,a+70,b-114,a-127,b+70,a+11,b-127,a-114 | b+11,a+9,b-72,a-93,b+9,a+11,b-93,a-72 | Yes |
| 5h9u | a-9,b-58,a-67,b-9,a-58,b-67 | b-2,a-5,b-85,a-2,b-5,a-85 | No |
| 3zq5 | a+77,b-92,a-150,b+77,a+151,b-150,a-92,b+151 | b+14,a+13,b-98,a-102,b+13,a+14,b-102,a-98 | No |
| 4ktv | b-2,a-157,b-158,a-2,b-157,a-158 | b-12,a-18,b-65,a-12,b-18,a-65 | Yes |
| 1uak | b-40,a-51,b-70,a-40,b-51,a-70 | b-36,a-37,b-44,a-36,b-37,a-44 | Yes |
| 1fug | a-6,b-14,a-86,b-6,a-14,b-86 | b-8,a-5,b-35,a-8,b-5,a-35 | No |
| 1p7l | b-5,a-7,b-15,a-5,b-7,a-15 | b-8,a-5,b-35,a-8,b-5,a-35 | Yes |
| 2etl | a+2,a+5,b+7,a+50,b+51,b+2,b-59,a+101,b+102, a+7,b+5,b-107,b+101,a+51,b+50,a+102,a-107,a-59 | b+1,a+3,b+14,a+21,b+22,a+1,b+3,a+22,b+21,a+14 | No |
| 2k0a | a+9,b+11,a+15,b+9,a+11,b+15 | a+8,b+9,a+15,b+8,a+9,b+15 | Yes |
| 1js1 | a-131,b-132,a-135,b-131,a-132,b-135 | b-112,a-119,b-126,a-112,b-119,a-126 | No |
| 1nxz | b-61,a-64,b-66,a-61,b-64,a-66 | a-94,b-95,a-100,b-94,a-95,b-100 | No |
| 4v1a | a+12,b+100,a+172,b+12,a+100,b+172 | b+20,a+51,b+111,a+20,b+51,a+111 | No |
| 4cng | a-14,b-43,a-44,b-14,a-43,b-44 | b-25,a-27,b-30,a-25,b-27,a-30 | No |
| 4l2z | b-11,a-51,b-56,a-11,b-51,a-56,b+100,a+116,b+127,a+100,b+116,a+127 | a-1,a+10,b-23,a-35,b+142,b-1,a+148,a-23,b-35,b+303, +305,b+314,a+303,b+305,a+314,b+148,b+10,a+142 | No |
| 4l4q | b-19,a-52,b-22,a-19,b-52,a-22 | b-32,a-206,b-227,a-32,b-206,a-227 | Yes |
| 4jkj | a+7,b+48,a+49,b+106,a+105,b+7,a+106,b+49,a+48,b+105 | b+1,a+3,b+14,a+21,b+22,a+1,b+3,a+22,b+21,a+14 | No |
| 4ndn | b-2,a-44,b-178,a-2,b-44,a-178 | b-12,a-18,b-65,a-12,b-18,a-65 | Yes |
| 1js1 | a-131,b-132,a-135,b-131,a-132,b-135 | b-112,a-119,b-126,a-112,b-119,a-126 | No |
| 1mxi | a-22,b-33,a-38,b-22,a-33,b-38 | a-43,b-47,a-51,b-43,a-47,b-51 | Yes |
| 1p7l | b-5,a-7,b-15,a-5,b-7,a-15 | b-8,a-5,b-35,a-8,b-5,a-35 | Yes |
| 1yrl | a+123,b-118,a-145,b+123,a+223,b-145,a-118,b+223 | a-212,b-242,a+240,b+249,a-242,b-212,a+249,b+240 | No |
| 2g7m | b-87,a-92,b-96,a-87,b-92,a-96 | b-112,a-119,b-126,a-112,b-119,a-126 | Yes |
| 2qmm | b-60,a-64,b-66,a-60,b-64,a-66 | a-75,b-76,a-77,b-75,a-76,b-77 | No |
| 2yy8 | b-8,a-42,b-43,a-8,b-42,a-43 | a-43,b-48,a-56,b-43,a-48,b-56 | No |
| 3bbd | b-89,a-88,b-97,a-89,b-88,a-97 | a-53,b-55,a-63,b-53,a-55,b-63 | No |
| 3c2w | a+23,b+18,a-112,b-134,a+18,b+23,a-134,b-112 | a+2,b+138,a-199,b-221,a+138,b+2,a-221,b-199 | Yes |
| 3dcm | b-33,a-38,b-42,a-33,b-38,a-42 | b-78,a-65,b-50,a-78,b-65,a-50 | Yes |
| 1p7l | b-5,a-7,b-15,a-5,b-7,a-15 | b-8,a-5,b-35,a-8,b-5,a-35 | Yes |
| 1x7o | a-92,b-95,a-103,b-92,a-95,b-103 | a-128,b-129,a-135,b-128,a-129,b-135 | Yes |
| 1znc | a-3,b-10,a-48,b-3,a-10,b-48 | b-11,a-14,b-16,a-11,b-14,a-16 | No |
| 2ha8 | a-25,b-36,a-41,b-25,a-36,b-41 | b-845,a-846,b-851,a-845,b-846,a-851 | No |
| 2obv | b-3,a-43,a-44,b-173,a-176,b-44,a+232,a-3,a-173,b-176,b-43,b+232 | a-36,b-160,a-200,b-36,a-160,b-200 | No |
| 2yfk | a-143,b-171,a-162,b-143,a-171,b-162 | Unkotted | No |
| 3c2w | a+23,b+18,a-112,b-134,a+18,b+23,a-134,b-112 | a+2,b+138,a-199,b-221,a+138,b+2,a-221,b-199 | Yes |
| 3g6o | a+15,b+14,a-54,b-81,a+14,b+15,a-81,b-54 | a+2,b+138,a-199,b-221,a+138,b+2,a-221,b-199 | Yes |
| 3m5d | a-96,b-97,a-104,b-96,a-97,b-104 | Unkotted | No |
| 3n4j | b-37,a-38,b-44,a-37,b-38,a-44 | a-4,b-37,a-40,b-4,a-37,b-40 | No |
| 3ulk | b+190,a-195,b-209,a+190,b+231,a-209,b-195,a+231 | a-212,b-242,a+240,b+249,a-242,b-212,a+249,b+240 | No |

# Applications of Knot Theory for the Improvement of the AlphaFold Protein Database

Table 3: Comparison of Alexander-Briggs Knots (Full Data)

| PDB ID | KnotProt Knot | AlphaFold Knot | Knot Match |
|---|---|---|---|
| 7qi6 | 3-1 | 3-1 | Yes |
| 6km1 | 3-1 | 3-1 | Yes |
| 6ptx | 4-1 | 4-1 | Yes |
| 1ipa | 3-1 | 3-1 | Yes |
| 6ux1 | 3-1 | 3-1 | Yes |
| 3hkn | | | Yes |
| 4twm | 3-1 | 3-1 | Yes |
| 5vik | 4-1 | 4-1 | Yes |
| 5h9u | 3-1 | 3-1 | Yes |
| 3zq5 | 4-1 | 4-1 | Yes |
| 4ktv | 3-1 | 3-1 | Yes |
| 1uak | 3-1 | 3-1 | Yes |
| 1fug | 3-1 | 3-1 | Yes |
| 1p7l | 3-1 | 3-1 | Yes |
| 2etl | 5-2 | 5-2 | Yes |
| 2k0a | 3-1 | 3-1 | Yes |
| 1js1 | 3-1 | 3-1 | Yes |
| 1nxz | 3-1 | 3-1 | Yes |
| 4v1a | 3-1 | 3-1 | Yes |
| 4cng | 3-1 | 3-1 | Yes |
| 4l2z | 3-1 | 3-1 | Yes |
| 4l4q | 3-1 | 3-1 | Yes |
| 4jkj | 5-2 | 5-2 | Yes |
| 4ndn | 3-1 | 3-1 | Yes |
| 1js1 | 3-1 | 3-1 | Yes |
| 1mxi | 3-1 | 3-1 | Yes |
| 1p7l | 3-1 | 3-1 | Yes |
| 1yrl | 4-1 | 4-1 | Yes |
| 2g7m | 3-1 | 3-1 | Yes |
| 2qmm | 3-1 | 3-1 | Yes |
| 2yy8 | 3-1 | 3-1 | Yes |
| 3bbd | 3-1 | 3-1 | Yes |
| 3c2w | 4-1 | 4-1 | Yes |
| 3dcm | 3-1 | 3-1 | Yes |
| 1p7l | 3-1 | 3-1 | Yes |
| 1x7o | 3-1 | 3-1 | Yes |
| 1znc | 3-1 | 3-1 | Yes |
| 2ha8 | 3-1 | 3-1 | Yes |
| 2obv | 3-1 | 3-1 | Yes |
| 2yfk | 3-1 | Unknotted | No |
| 3c2w | 4-1 | 4-1 | Yes |
| 3g6o | 4-1 | 4-1 | Yes |
| 3m5d | 3-1 | Unknotted | No |
| 3n4j | 3-1 | 3-1 | Yes |
| 3ulk | 4-1 | 4-1 | Yes |